\documentclass[aps,prl,twocolumn,groupedaddress]{revtex4}
\usepackage{longtable}
\usepackage{graphics}
\usepackage{graphicx,epsfig,pstcol,pstricks}

\begin{document}

%\preprint{}

\title{Hard biaxial ellipsoids revisited: numerical results}

\author{Carl McBride}
\author{Enrique Lomba}
%\thanks{}
\affiliation{Instituto de Qu\'{\i}mica F\'{\i}sica Rocasolano (CSIC), Serrano 119, 28006 Madrid, Spain}


\date{November 14, 2006}

\begin{abstract}
Monte Carlo simulations are performed for hard ellipsoids for a number of 
values of its semi-axes in the range  $c/a \in \{0.1,10\}$. 
The isotropic phase  results are compared to the Vega equation of state
[Mol. Phys. {\bf 92} 651 (1997)].
The position of the isotropic-nematic transition is also evaluated.
The biaxial phase is seen to form only after the previous  formation of 
a discotic phase.
\end{abstract}

%\keywords{}

\maketitle

\section{Introduction}
%%%%%%%%%%%%%%%%%%%%%%%%%%%%%%%%%%%%%%%%%%%%%%%%%%%%%%%%%%%%%%%%%%%
One of the first ever models used in computer simulation studies of the 
fluid phase  was the hard sphere. 
The next most obvious choice of model is the hard ellipsoid; an affine transformation of the 
hard sphere. The orientation of the model now becomes a variable within the simulation. 
Having a model that incorporates orientation facilitates the study of 
orientationally ordered phases, such as those associated with 
liquid crystals.

As with hard spheres \cite{JCP_1953_21_01087,JCP_1957_27_01208,LASLR_1963_2827,JCP_1970_52_00729} the first 
simulations of hard ellipsoids were performed for two-dimensional systems \cite{JCP_1972_56_04729,PRA_1990_42_002126}.
Hard spheres are only capable of forming two phases; the fluid and the solid. There is no
`gas-liquid' transition, due to the lack of attractive forces. 
However, the hard ellipsoid system also has a plastic crystal (for small axis ratios) and a nematic phase.
Prolate and prolate-like biaxial models, {\it i.e.} models of the form $a \leq  b < c$ with $b <  \sqrt{ac}$ where $a$, $b$
and $c$ are the semi-axes of the ellipsoid, are able to form a uniaxial nematic phase, often denominated
as $N_+$.
Oblate and oblate-like biaxial models, with  $a <  b \leq c$ and $b >  \sqrt{ac}$, can form a uniaxial `discotic' phase ($N_-$).
A tentative phase diagram for uniaxial ellipsoids was proposed by Frenkel and co-workers 
\cite{PRL_1984_52_000287,MP_2002_100_0199,MP_1985_55_1171,MP_1985_55_1193}.
As well as these nematic and discotic phases, Freiser \cite{PRL_1970_24_001041} predicted
that `long-flat' molecules could form a biaxial phase $N_B$, which has recently been discovered experimentally
\cite{PRL_2004_92_145505,PRL_2004_92_145506}.

The study of hard bodies provide reference systems for use in perturbation theories which add long range attractive interactions \cite{MP_1988_63_1103}.
They can also form the monomer units of larger molecules \cite{JCP_2002_117_10370,FPE_2002_194_0227}.
Hard ellipsoids continue to be of interest \cite{PRE_2006_74_031124,JCP_2006_124_104509}, 
having recently been shown that a maximally random jammed (MRJ) packing fraction of $\phi=0.7707$ is possible for
models whose maximal aspect ratio is greater than $\sqrt{3}$.
\cite{S_2004_303_00990,PRL_2004_92_255506},
Such high packing fractions were 
obtained using an `event-driven' molecular dynamics code \cite{JCompP_2005_202_0737,JCompP_2005_202_0765}
and were confirmed experimentally using latex particles and the like \cite{L_2006_22_06605,PRL_2005_94_198001}.

Biaxial ellipsoids have, however, received comparatively little attention, simulations having been performed
principally by Allen \cite{LC_1990_8_0499} and by Camp and Allen \cite{JCP_1997_106_06681}.
Extensive simulation results have been presented previously for uniaxial hard ellipsoids,
notably those of Frenkel and Mulder for $c/a$ in the range $[1/3,3]$ \cite{MP_1985_55_1171}
In this publication a number of simulations are performed for biaxial ellipsoids within the region 
 $c/a$  for  $[0.1,10]$. Also in this work the isotropic equation of state is compared to theory.
%%%%%%%%%%%%%%%%%%%%%%%%%%%%%%%%%%%%%%%%%%%%%%%%%%%%%%%%%%%%%%%%%%%
\section{Simulation technique}
%%%%%%%%%%%%%%%%%%%%%%%%%%%%%%%%%%%%%%%%%%%%%%%%%%%%%%%%%%%%%%%%%%%
Standard Metropolis Monte Carlo sampling was used \cite{JCP_1953_21_01087}.
The decision to reject or accept a trial move for hard bodies 
is based simply on whether two bodies overlap or not. 
For hard spheres ($a=b=c$) this criteria is trivial; if the distance between 
two bodies is less than twice the radius then they overlap.
For hard ellipsoids the situation is more complicated, having to take into account
the orientations of the ellipsoids as well as the distance between them.
One of the first overlap criteria was proposed for two-dimensional ellipsoids 
by Vieillard-Baron \cite{JCP_1972_56_04729,CAGD_2001_18_531}. 
Perram and Wertheim produced an
overlap algorithm for three-dimensional ellipsoids \cite{JCompP_1985_58_0409}.
In this study the Perram-Wertheim criteria was used.

All of the Monte Carlo simulations were performed in the $NpT$ ensemble (with $k_BT=1$), having cubic 
boundary conditions, with the system comprising of $N=343$ hard ellipsoids. 
The runs consisted of between 75 and 150  kilocycles for equilibration followed by a further
75-150  kilocycles for the production of thermodynamic data. 
Each simulation was initiated from the 
final configuration of the previous, lower pressure, run.
Throughout the simulations the uniaxial order parameter $S_2$ was monitored 
for each of the three axis 
by calculating 
%(here for the $c$ axis)
\begin{equation}
%S_2 = \left< \bf {Z} \cdot \bf {Q^{ZZ}}   \cdot \bf {Z} \right> = \left< \frac{1}{2}(3 \cos^2 \theta -1) \right>.
S_2 =  \left< \frac{1}{2}(3 \cos^2 \theta_i -1) \right>,
\end{equation}
where $\theta_i$ is the angle between ellipsoid $i$ and the director vector
\cite{PRA_1974_10_001881,MP_1984_52_1303,chapterMPLC_1979_ch3_p051}.

The prolate parameters are typified by $a=b$, the oblate parameters by 
$b=c$ and the biaxial parameters by $a \neq b \neq c$ where in this work
$a < b < c$ (in this work we define $a=1$).

Boublik and Nezbeda have shown \cite{CCCC_1986_51_2301_photocopy} 
that hard convex bodies can be described using three geometric expressions;
the volume, $V$, the surface area, $S$ and the mean radius of curvature, $R$.
The second virial coefficient, $B_2$ for any hard convex body is given by the 
Isihara-Hadwiger formula \cite{JCP_1950_18_01446,JPSJ_1951_06_00040},
and provides a measure of the excluded volume of a `molecule' due 
to the presence of a second molecule,
\begin{equation}
B_2 = RS+V,
\end{equation}
or 
\begin{equation}
\frac{B_2}{V}=1+3\alpha,
\end{equation}
if one defines the `non-sphericity' parameter \cite{JPSJ_1951_06_00046} %check this
\begin{equation}
\alpha = \frac{RS}{3V}.
\end{equation}
Values for  $R$, $S$ and $V$  for ellipsoids are provided in Appendix A.
%%%%%%%%%%%%%%%%%%%%%%%%%%%%%%%%%%%%%%%%%%%%%%%%%%%%%%%%%%%%%%%%%%%
%%%%%%%%%%%%%%%%%%%%%%%%%%%%%%%%%%%%%%%%%%%%%%%%%%%%%%%%%%%%%%%%%%%
\section{Results}
%%%%%%%%%%%%%%%%%%%%%%%%%%%%%%%%%%%%%%%%%%%%%%%%%%%%%%%%%%%%%%%%%%%
%%%%%%%%%%%%%%%%%%%%%%%%%%%%%%%%%%%%%%%%%%%%%%%%%%%%%%%%%%%%%%%%%%%
\subsection{Isotropic phase equation of state} 
%%%%%%%%%%%%%%%%%%%%%%%%%%%%%%%%%%%%%%%%%%%%%%%%%%%%%%%%%%%%%%%%%%%
Accurate equations of state (EOS) are necessary if the hard ellipsoid fluid
is to be used as a reference system in perturbation theories.
A number of equations of state have been proposed over the years 
to reproduce the behaviour of the isotropic phase of the ellipsoid
system; notably those of Nezbeda \cite{CPL_1976_41_0055},
Parsons \cite{PRA_1979_19_001225}, and Song and Mason \cite{PRA_1990_41_003121}.
It should be noted that all of these equations of state implicitly assume that 
the EOS is symmetric with respect to prolate/oblate models, since  they are built either 
directly using $B_2$, or indirectly using $\alpha$. This is because $B_2 (1\times1\times x) = B_2 (1\times x \times x)$.

In order to reproduce simulation results, higher virial coefficients are
required,
where it has been shown that the prolate/oblate symmetry 
breaks down starting from $B_3$ \cite{MP_1989_66_1261,MP_2001_99_0187}, 
as used in the uniaxial EOS of Maeso and Solana \cite{MP_1993_79_1365} 
or the biaxial EOS of Vega \cite{MP_1997_92_0651}.
In this work the simulation results are compared to the biaxial Vega equation of state.
The Vega EOS is given by 
\begin{eqnarray}
Z = &1&+B_2^*y + B_3^*y^2 + B_4^*y^3 + B_5^*y^4 \nonumber \\
    &+& \frac{B_2}{4} \left( \frac{1+y+y^2-y^3}{(1-y)^3} \right. \nonumber\\ 
    &-&1 \left. -4y -10y^2 -18.3648y^3 - 28.2245y^4 \right),
\end{eqnarray}
where $Z$ is the compressibility factor and $y$ is the volume fraction, given by
$y= \rho V$ where $\rho$ is the number density.
The virial coefficients are given by the fits
\begin{eqnarray}
B_3^* &= & 10 + 13.094756 \alpha'  - 2.073909\tau' + 4.096689 \alpha'^2 \nonumber \\
        &+&  2.325342\tau'^2 - 5.791266\alpha' \tau',
\end{eqnarray}
\begin{eqnarray}
B_4^* &= &18.3648 + 27.714434\alpha' - 10.2046\tau' +  11.142963\alpha'^2 \nonumber \\
        &+& 8.634491\tau'^2 - 28.279451\alpha' \tau' \nonumber \\
        &-&  17.190946\alpha'^2 \tau' + 24.188979\alpha' \tau'^2 \nonumber \\
        &+& 0.74674\alpha'^3 - 9.455150\tau'^3,
\end{eqnarray}
and
\begin{eqnarray}
B_5^* &= &28.2245 + 21.288105\alpha' + 4.525788\tau' +  36.032793\alpha'^2 \nonumber \\
        &+& 59.0098\tau'^2 - 118.407497\alpha' \tau' \nonumber \\
        &+&  24.164622\alpha'^2 \tau' + 139.766174\alpha' \tau'^2 \nonumber \\
        &-& 50.490244\alpha'^3 - 120.995139\tau'^3 + 12.624655\alpha'^3\tau',
\end{eqnarray}
where
\begin{equation}
\tau' = \frac{4 \pi R^2}{S} -1,
\end{equation}
and
\begin{equation}
\alpha' = \frac{RS}{3V}-1.
\end{equation}
Values for $R$ and $S$ for the models studied in the work are given in Appendix A.

\begin{figure}
 \includegraphics[height=200pt,width=250pt]{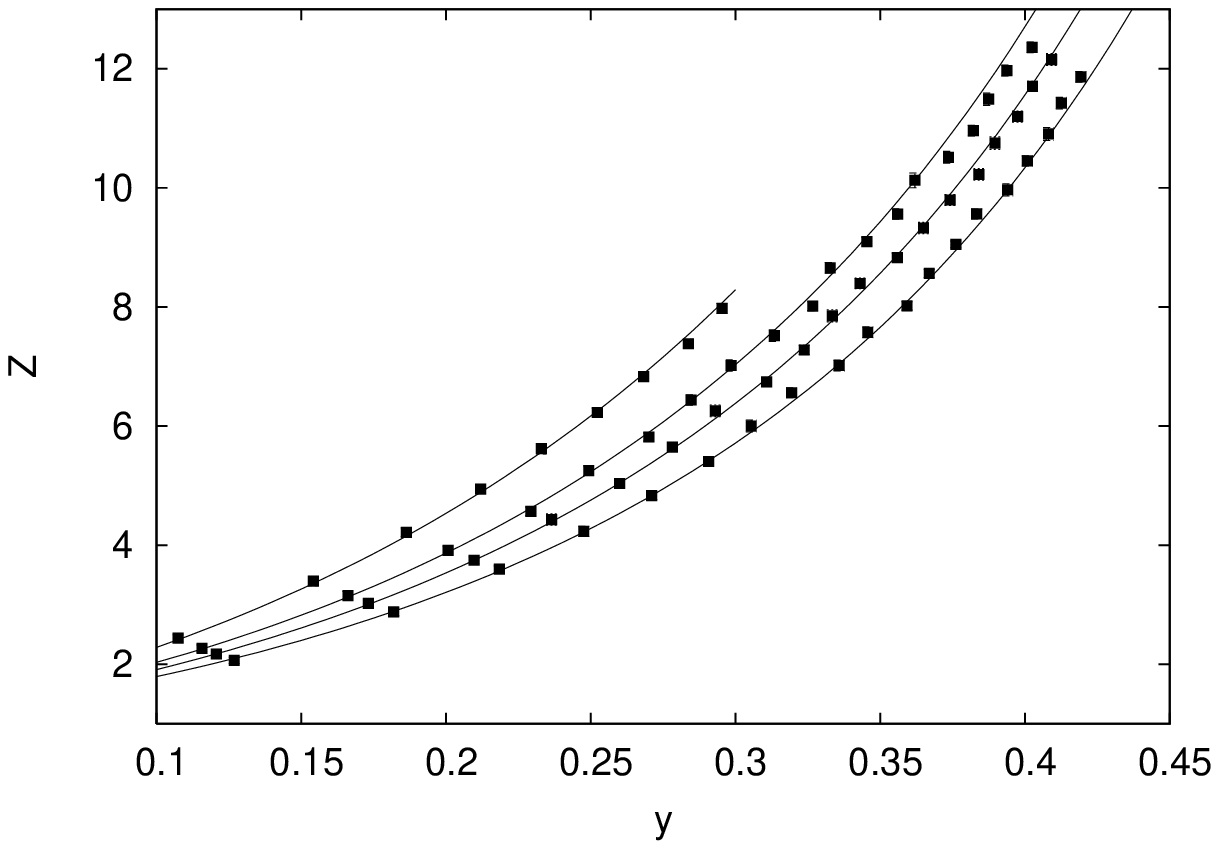}%
 \caption{\label{graph_c} Plot of the simulation results for the biaxial molecules $1\times2\times4$, $1\times2\times5$,
$1\times2\times6$ and the isotropic state points for $1\times2\times8$. The lines are the Vega equation of state.
}
 \end{figure}

\begin{figure}
 \includegraphics[height=200pt,width=250pt]{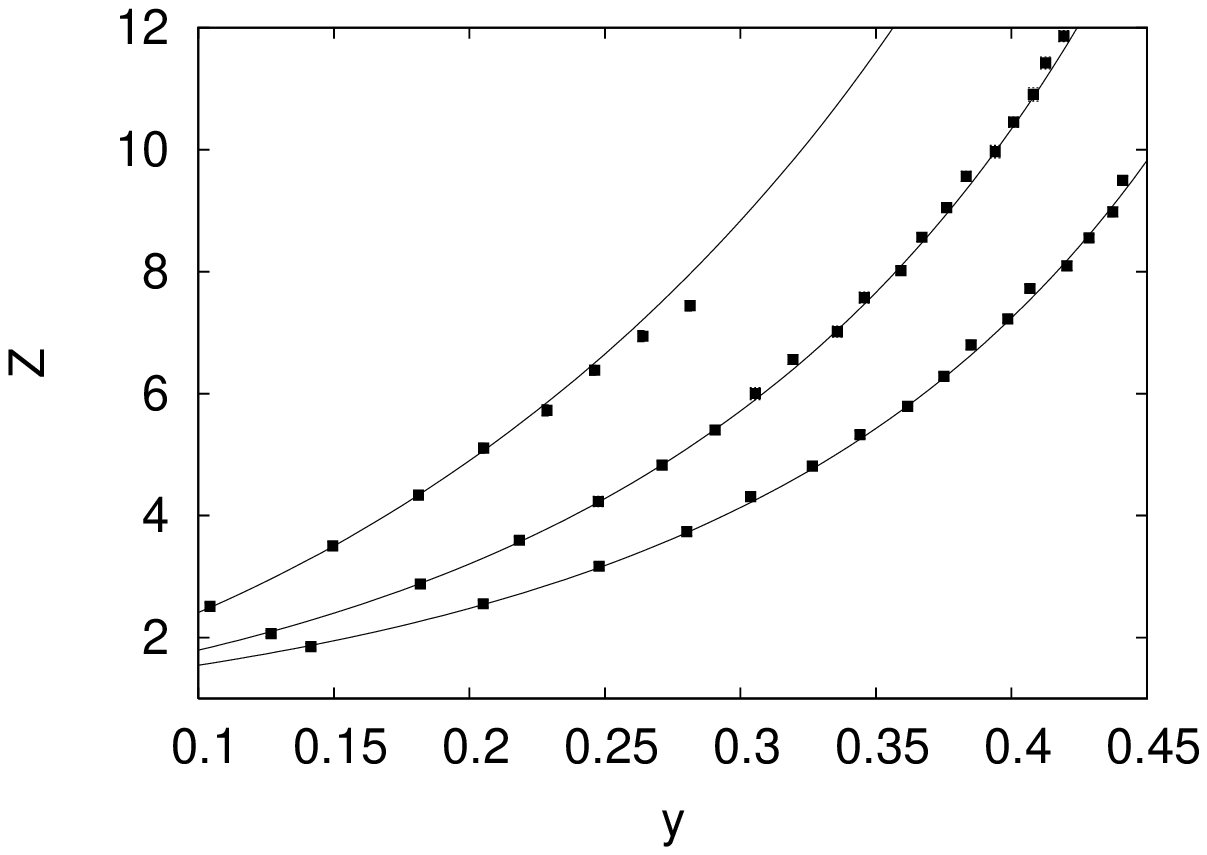}%  
 \caption{\label{figure_dual} Plot of the simulation results for the self-dual molecules $1\times1.25\times1.5625$, $1\times2\times4$ and the isotropic state points for $1\times3\times9$.
The lines are the Vega equation of state.
}
 \end{figure}

The effect of varying $c$ for a biaxial model is shown in Fig. \ref{graph_c}, where
various models from Table \ref{biaxial_table} are plotted.
\begin{turnpage}
 \begin{table*}%[H] add [H] placement to break table across pages
 \caption{\label{biaxial_table} Equations of state for biaxial ($a=1$, $a < b < c$) hard ellipsoids. Error in $y$ is $O(10^{-3})$, error in $Z$ is  $O(10^{-2})$}
 \begin{ruledtabular}
 ~~~~$b=2, c=5$ ~~~~~~~~$b=2, c=6$ ~~~~~~~~$b=2, c=8$ ~~~~~~~$b=3, c=6$ ~~~~~$b=3, c=8$ ~~~~~~$b=3, c=10$ ~~~~~~$b=5, c=8$ ~~~~~~$b=5, c=10$ ~~~~~~$b=8, c=10$\\
\begin{tabular}{|l|rr|rr|rr|rr|rr|rr|rr|rr|rr|}
\hline
$p^*$      &            y  &          Z &         y  &       Z &          y  &            Z &       y  &          Z  &     y &            Z &        y &          Z &        y &           Z &   y &     Z &   y     &  Z\\
\hline
0.5&0.120&2.17&0.115&2.26&0.107&2.43&0.114&2.28&0.107&2.44&0.102&2.56&0.103&2.52&0.097&2.67&0.093&2.80\\
1.0&0.173&3.02&0.166&3.15&0.154&3.39&0.165&3.16&0.153&3.40&0.145&3.60&0.146&3.56&0.141&3.70&0.134&3.89\\
1.5&0.209&3.74&0.200&3.91&0.186&4.21&0.198&3.96&0.196&4.21&0.176&4.45&0.180&4.34&0.170&4.59&0.165&4.74\\
2.0&0.236&4.42&0.229&4.56&0.211&4.94&0.227&4.60&0.212&4.93&0.200&5.21&0.206&5.06&0.199&5.25&0.194&5.39\\
2.5&0.259&5.03&0.249&5.25&0.232&5.62&0.246&5.30&0.233&5.60&0.224&5.82&0.226&5.77&0.220&5.93&0.219&5.96\\
3.0&0.278&5.64&0.270&5.81&0.252&6.22&0.269&5.82&0.253&6.20&0.243&6.45&0.249&6.30&0.243&6.44&0.249&6.29\\
3.5&0.293&6.25&0.284&6.43&0.268&6.83&0.283&6.45&0.272&6.72&0.258&7.08&0.270&6.76&0.269&6.80&0.286&6.40\\
4.0&0.310&6.74&0.298&7.01&0.283&7.38&0.297&7.03&0.284&7.36&0.280&7.48&0.288&7.25&0.301&6.95&0.308&6.78\\
4.5&0.323&7.27&0.313&7.51&0.295&7.97&0.310&7.59&0.302&7.77&0.292&8.05&0.313&7.51&0.320&7.34&0.327&7.19\\
5.0&0.333&7.85&0.326&8.01&0.310&8.43&0.324&8.06&0.314&8.33&0.310&8.43&0.338&7.73&0.334&7.83&0.341&7.67\\
5.5&0.343&8.39&0.332&8.65&0.323&8.90&0.333&8.63&0.330&8.70&0.328&8.77&0.352&8.17&0.349&8.23&0.357&8.05\\
6.0&0.355&8.82&0.345&9.09&0.342&9.16&0.345&9.09&0.338&9.28&0.345&9.08&0.363&8.63&0.361&8.68&0.365&8.58\\
6.5&0.364&9.32&0.356&9.56&0.357&9.52&0.355&9.58&0.363&9.37&0.355&9.58&0.372&9.13&0.371&9.17&0.377&9.01\\
7.0&0.374&9.79&0.361&10.12&0.370&9.88&0.362&10.11&0.367&9.98&0.366&9.99&0.388&9.43&0.381&9.60&0.391&9.36\\
7.5&0.384&10.22&0.373&10.51&0.390&10.05&0.370&10.60&0.379&10.33&0.378&10.37&0.397&9.88&0.391&10.04&0.398&9.84\\
8.0&0.389&10.75&0.382&10.96&0.395&10.59&0.385&10.87&0.387&10.80&0.389&10.76&0.410&10.20&0.399&10.48&0.409&10.24\\
8.5&0.397&11.19&0.387&11.48&0.399&11.13&0.394&11.27&0.399&11.13&0.392&11.33&0.414&10.72&0.411&10.81&0.419&10.62\\
9.0&0.402&11.70&0.393&11.96&0.415&11.35&0.402&11.70&0.406&11.59&0.395&11.90&0.421&11.17&0.415&11.34&0.425&11.06\\
9.5&0.409&12.15&0.402&12.35&0.421&11.78&0.409&12.15&0.414&12.00&0.408&12.17&0.432&11.49&0.420&11.82&0.432&11.50\\
\hline
 \end{tabular}
 \end{ruledtabular}
 \end{table*}
\end{turnpage}
As can be seen, the Vega EOS provides an excellent fit for all of the isotropic state points.
This is also demonstrated in a plot  of the self-dual models, where $b=\sqrt{ac}$ (Fig. \ref{figure_dual})
taken from Table \ref{self_dual_table}.
 \begin{table}%[H] add [H] placement to break table across pages
 \caption{\label{self_dual_table} Equations of state for biaxial self-dual ($b=\sqrt{ac}$) hard ellipsoids. Error in $y$ is $O(10^{-3})$, error in $Z$ is  $O(10^{-2})$}
 \begin{ruledtabular}
$b=1.25, c=1.5625$ ~~~~~~~$b=2, c=4$ ~~~~~~~$b=3, c=9$\\
\begin{tabular}{|l|rr|rr|rr|}
\hline
$p^*$     &            y  &          Z &         y  &       Z  &         y  &       Z \\
\hline
0.5  &0.141      & 1.85    & 0.126     & 2.06   &  0.104  &  2.51  \\
1.0  &0.205      & 2.55    & 0.181     & 2.87    & 0.149  &  3.50   \\
1.5  &0.247      &3.17     & 0.218     & 3.59    & 0.181  &  4.33   \\
2.0  &0.280      & 3.73    & 0.247     & 4.23   &  0.205  &  5.10    \\
2.5  &0.303      & 4.31    & 0.271     & 4.82   &  0.228  &  5.76    \\
3.0  &0.326      & 4.81    & 0.290     & 5.40   &  0.246  &  6.38    \\
3.5  &0.344      & 5.32    & 0.305     & 6.00 &    0.264  &  6.94      \\
4.0  &0.361      & 5.79    & 0.319     & 6.55    & 0.281  &  7.44   \\
4.5  &0.375      &6.28     & 0.335     & 7.01    & 0.301  &  7.82   \\
5.0  &0.385      & 6.79    & 0.345     & 7.57    & 0.315  &  8.29   \\
5.5  &0.389      & 7.22    & 0.359     & 8.01   &  0.332  &  8.67    \\
6.0  &0.406      &7.72     & 0.366     & 8.56    &  0.344  & 9.12   \\
6.5  &0.420      &8.09     & 0.376     & 9.04    &  0.356  &  9.54   \\
7.0  &0.428      &8.55     & 0.383     & 9.56   &  0.364  &  10.06     \\
7.5  &0.437      & 8.97    & 0.394     & 9.96      &  0.376  &  10.41      \\
8.0  &0.441      &9.49     & 0.400     & 10.45   &  0.384  &  10.88     \\
8.5  & 0.451     &9.86     & 0.408     & 10.90    &  0.395  &  11.26     \\
9.0  &0.458      & 10.28   & 0.412     & 11.42     &  0.406  &  11.58     \\
9.5  &0.463      & 10.72   & 0.419     & 11.86     &  0.410  &  12.11     \\
\hline
 \end{tabular}
 \end{ruledtabular}
 \end{table}
 \begin{figure}
 \includegraphics[height=200pt,width=250pt]{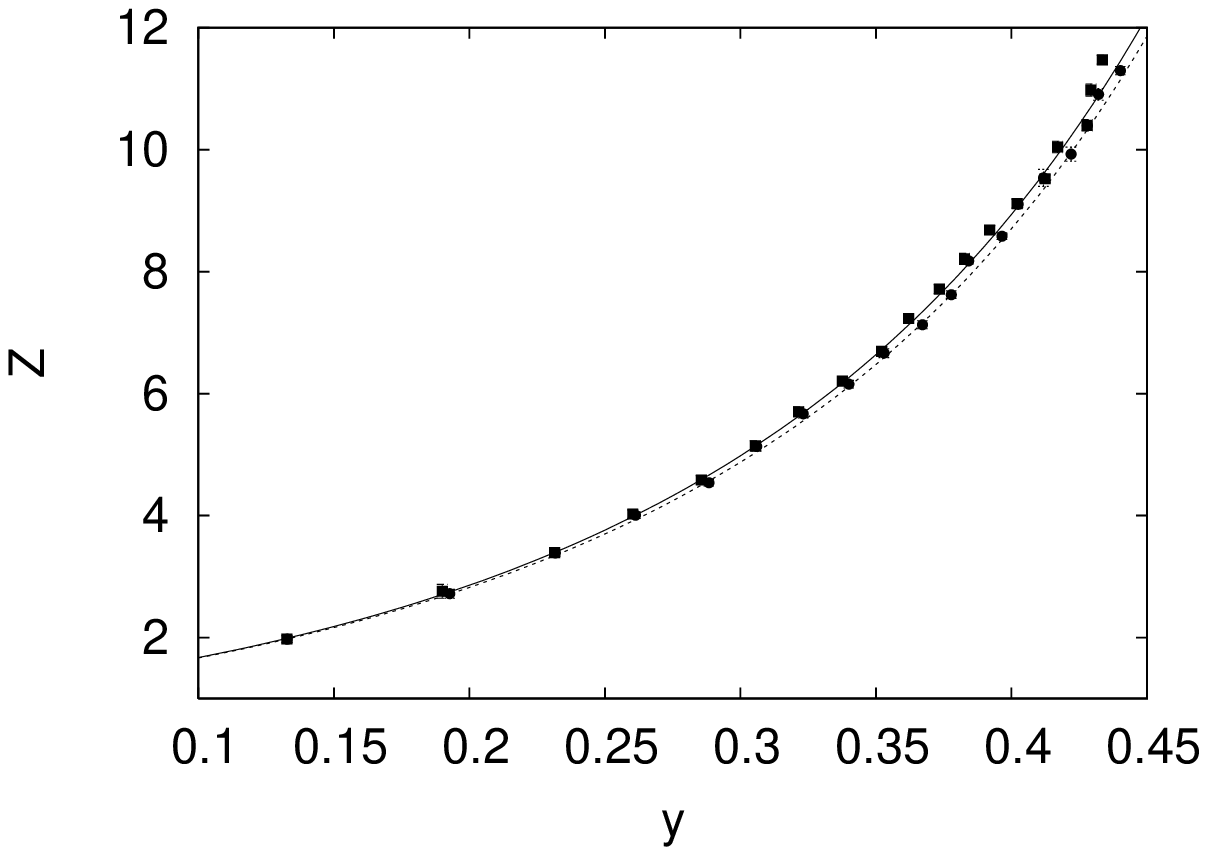}%
 \caption{\label{fig_2.5} Plot of the  results for the oblate $1\times2.5\times2.5$ model (black squares) and the
prolate $1\times1\times2.5$ model (black circles) along with the Vega EOS (oblate: solid line, prolate: dashed line)}
 \end{figure}
 \begin{figure}
 \includegraphics[height=200pt,width=250pt]{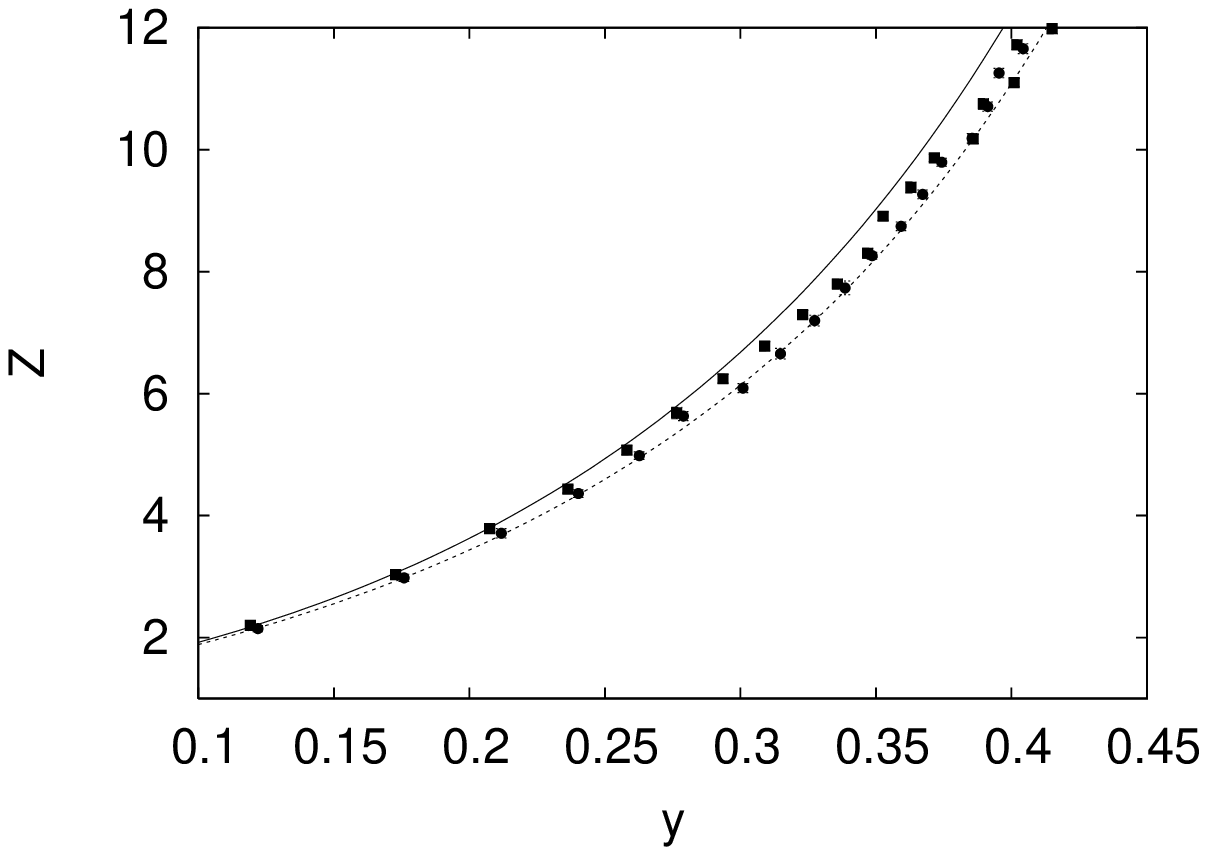}%
 \caption{\label{fig_4} Plot of the  results for the oblate $1\times4\times4$ model (black squares) and the
prolate $1\times1\times4$ model (black circles) along with the Vega EOS (oblate: solid line, prolate: dashed line)}
 \end{figure}
In figures \ref{fig_2.5} and \ref{fig_4}
the isotropic equations of state are plotted for the $1\times 2.5 \times 2.5$, $1\times 1 \times 2.5$
and for the $1\times 4 \times 4$, and $1\times 1 \times 4$ models.
It can be seen that for these modest anisotropies the oblate/prolate curves are indeed almost coincident, as 
assumed in the simple EOS built using $B_2$ only.
%Plots (Figs. 3-8)are also provided for various $1\times x \times x$, $1\times 1 \times x$ oblate/prolate models, where $x \in \{2.5, 4, 5,6,8,10\}$.
The numerical results are presented in Tables \ref{oblate_table} and \ref{prolate_table}.
 \begin{table*}%[H] add [H] placement to break table across pages
 \caption{\label{oblate_table} Equations of state for oblate ($a=1$, $b=c$) hard ellipsoids. Error in $y$ is $O(10^{-3})$, error in $Z$ is  $O(10^{-2})$}
 \begin{ruledtabular}
c ~~~~~~~~ 2.5 ~~~~~~~~~~~~~~~~ 4 ~~~~~~~~~~~~~~~~ 5  ~~~~~~~~~~~~~~~~~~~ 6 ~~~~~~~~~~~~~~~~~~~~~~~~~~~ 8 ~~~~~~~~~~~~~~~~~~~ 10 \\
\begin{tabular}{|l|rr|rr|rr|rr|rr|rr|}
\hline
$p^*$        & y  &       Z &        y  &     Z &       y  &       Z &      y  &      Z &        y  &       Z  &        y  &       Z\\
\hline
0.5        & 0.132  & 1.97  & 0.119  & 2.19 & 0.113   &      2.31      & 0.108  & 2.41 & 0.097  & 2.70 & 0.090  & 2.90  \\
1.0        & 0.189  & 2.76  & 0.172  & 3.03 & 0.162   &      3.23      & 0.153  & 3.40 & 0.140  & 3.71 & 0.129  & 4.03  \\
1.5        & 0.231  & 3.39  & 0.207  & 3.78 & 0.196   &      4.00      & 0.185  & 4.24 & 0.171  & 4.57 & 0.161  & 4.85  \\
2.0        & 0.260  & 4.02  & 0.236  & 4.43 & 0.220   &      4.75      & 0.213  & 4.91 & 0.195  & 5.34 & 0.192  & 5.45  \\
2.5        & 0.285  & 4.58  & 0.258  & 5.07 & 0.243   &      5.38      & 0.233  & 5.61 & 0.226  & 5.76 & 0.240  & 5.44  \\
3.0        & 0.305  & 5.14  & 0.276  & 5.68 & 0.263   &      5.96      & 0.252  & 6.21 & 0.257  & 6.10 & 0.270  & 5.80  \\
3.5        & 0.321  & 5.70  & 0.293  & 6.24 & 0.282   &      6.49      & 0.273  & 6.69 & 0.281  & 6.51 & 0.293  & 6.24  \\
4.0        & 0.337  & 6.20  & 0.308  & 6.77 & 0.293   &      7.13      & 0.292  & 7.15 & 0.307  & 6.82 & 0.310  & 6.73  \\
4.5        & 0.352  & 6.69  & 0.323  & 7.29 & 0.312   &      7.52      & 0.306  & 7.69 & 0.324  & 7.26 & 0.326  & 7.22  \\
5.0        & 0.362  & 7.23  & 0.335  & 7.79 & 0.325   &      8.05      & 0.327  & 8.00 & 0.341  & 7.65 & 0.345  & 7.57  \\
5.5        & 0.373  & 7.71  & 0.346  & 8.30 & 0.335   &      8.58      & 0.336  & 8.56 & 0.351  & 8.19 & 0.356  & 8.08  \\
6.0        & 0.382  & 8.21  & 0.352  & 8.90 & 0.348   &      9.01      & 0.355  & 8.82 & 0.369  & 8.50 & 0.371  & 8.46  \\
6.5        & 0.391  & 8.68  & 0.362  & 9.38 & 0.358   &      9.49      & 0.375  & 9.07 & 0.374  & 9.09 & 0.379  & 8.96  \\
7.0        & 0.402  & 9.11  & 0.371  & 9.86 & 0.374   &      9.78      & 0.383  & 9.54 & 0.394  & 9.27 & 0.389  & 9.42  \\
7.5        & 0.412  & 9.52  & 0.385  & 10.17 & 0.386  &      10.16      & 0.394  & 9.95 & 0.404  & 9.70 &  0.399 & 9.83  \\ 
8.0        & 0.417  & 10.04 & 0.389  & 10.75 & 0.398  &      10.49      & 0.404  & 10.34 & 0.413  & 10.13 & 0.404  & 10.34  \\
8.5        & 0.427  & 10.40 & 0.400  & 11.09 & 0.415  &      10.72      & 0.415  & 10.70 & 0.416  & 10.68 & 0.415  & 10.72  \\
9.0        & 0.429  & 10.97 & 0.402  & 11.72 & 0.422  &      11.14      & 0.424  & 11.09 & 0.423  & 11.12 & 0.421  & 11.18  \\
9.5        & 0.433  & 11.47 & 0.415  & 11.98 & 0.437  &      11.36      & 0.431  & 11.53 & 0.428  & 11.60 & 0.430  & 11.56  \\
 \end{tabular}
 \end{ruledtabular}
 \end{table*}
 \begin{table*}%[H] add [H] placement to break table across pages
 \caption{\label{prolate_table} Equations of state for prolate ($a=b=1$) hard ellipsoids. Error in $y$ is $O(10^{-3})$, error in $Z$ is  $O(10^{-2})$}
 \begin{ruledtabular}
c ~~~~~~~~ 2.5 ~~~~~~~~~~~~~~~~ 4 ~~~~~~~~~~~~~~~~ 5  ~~~~~~~~~~~~~~~~~~~ 6 ~~~~~~~~~~~~~~~~~~~~~~~~~~~ 8 ~~~~~~~~~~~~~~~~~~~ 10 \\
\begin{tabular}{|l|rr|rr|rr|rr|rr|rr|}
\hline
$p^*$        & y  &       Z &        y  &     Z &       y  &       Z &      y  &      Z &        y  &       Z  &        y  &       Z\\
\hline
0.5        & 0.132  & 1.97 & 0.121  & 2.14 & 0.115    &     2.26      & 0.110  & 2.37 &      0.101  & 2.56   & 0.095  & 2.75 \\
1.0        & 0.192  & 2.71 & 0.175  & 2.97 & 0.166    &     3.14      & 0.158  & 3.29 &      0.145  & 3.59   & 0.136  & 3.84  \\
1.5        & 0.231  & 3.38 & 0.211  & 3.70 & 0.200    &     3.92      & 0.191  & 4.09 &      0.178  & 4.41   & 0.166  & 4.71  \\
2.0        & 0.261  & 4.00 & 0.240  & 4.36 & 0.226    &     4.61      & 0.218  & 4.80 &      0.201  & 5.19   & 0.192  & 5.43  \\
2.5        & 0.288  & 4.53 & 0.262  & 4.98 & 0.250    &     5.22      & 0.238  & 5.48 &      0.222  & 5.88   & 0.215  & 6.06  \\
3.0        & 0.305  & 5.13 & 0.278  & 5.63 & 0.268    &     5.86      & 0.256  & 6.13 &      0.240  & 6.54   & 0.228  & 6.88  \\
3.5        & 0.323  & 5.67 & 0.300  & 6.09 & 0.282    &     6.48      & 0.277  & 6.59 &      0.255  & 7.17   & 0.243  & 7.53  \\
4.0        & 0.340  & 6.15 & 0.314  & 6.65 & 0.300    &     6.96      & 0.289  & 7.23 &      0.269  & 7.78   & 0.258  & 8.11  \\
4.5        & 0.353  & 6.66 & 0.327  & 7.19 & 0.314    &     7.49      & 0.304  & 7.74 &      0.285  & 8.25   & 0.269  & 8.74  \\
5.0        & 0.367  & 7.12 & 0.338  & 7.73 & 0.325    &     8.03      & 0.311  & 8.40 &      0.298  & 8.77   & 0.285  & 9.15  \\
5.5        & 0.377  & 7.62 & 0.348  & 8.26 & 0.332    &     8.67      & 0.326  & 8.82 &      0.311  & 9.24   & 0.297  & 9.68  \\
6.0        & 0.384  & 8.17 & 0.359  & 8.74 & 0.346    &     9.06      & 0.334  & 9.39 &      0.321  & 9.76   & 0.308  & 10.19  \\
6.5        & 0.396  & 8.58 & 0.367  & 9.26 & 0.352    &     9.66      & 0.349  & 9.72 &      0.330  & 10.30  & 0.317  & 10.71  \\
7.0        & 0.402  & 9.10 & 0.374  & 9.79 & 0.365    &     10.03     & 0.356  & 10.27 &     0.349  & 10.49  & 0.327  & 11.17  \\
7.5        & 0.411  & 9.53 &  0.385  & 10.18 & 0.373   &     10.50     & 0.377  & 10.40 &     0.356  & 11.02  & 0.334  & 11.75  \\
8.0        & 0.422  & 9.92 &  0.391  & 10.70 & 0.375   &     11.16     & 0.389  & 10.76 &     0.367  & 11.40  & 0.337  & 12.40 \\
8.5        & 0.428  & 10.39 &0.395   &11.25 & 0.388   &     11.46     & 0.402  & 11.06 &     0.376  & 11.83  & 0.345  & 12.86  \\
9.0        & 0.432  & 10.90 &0.404  & 11.65 & 0.394   &     11.94     & 0.407  & 11.55 &     0.385  & 12.21  & 0.356  & 13.20  \\
9.5        & 0.440  & 11.29 &0.413  & 12.01 & 0.403   &     12.32     & 0.417  & 11.90 &     0.389  & 12.75  & 0.367  & 13.54  \\
 \end{tabular}
 \end{ruledtabular}
 \end{table*}
%%%%%%%%%%%%%%%%%%%%%%%%%%%%%%%%%%%%%%%%%%%%%%%%%%%%%%%%%%%%%%%%%%%
\subsection{ Isotropic-liquid crystal transition }
%%%%%%%%%%%%%%%%%%%%%%%%%%%%%%%%%%%%%%%%%%%%%%%%%%%%%%%%%%%%%%%%%%%
 \begin{figure}
 \includegraphics[height=200pt,width=250pt]{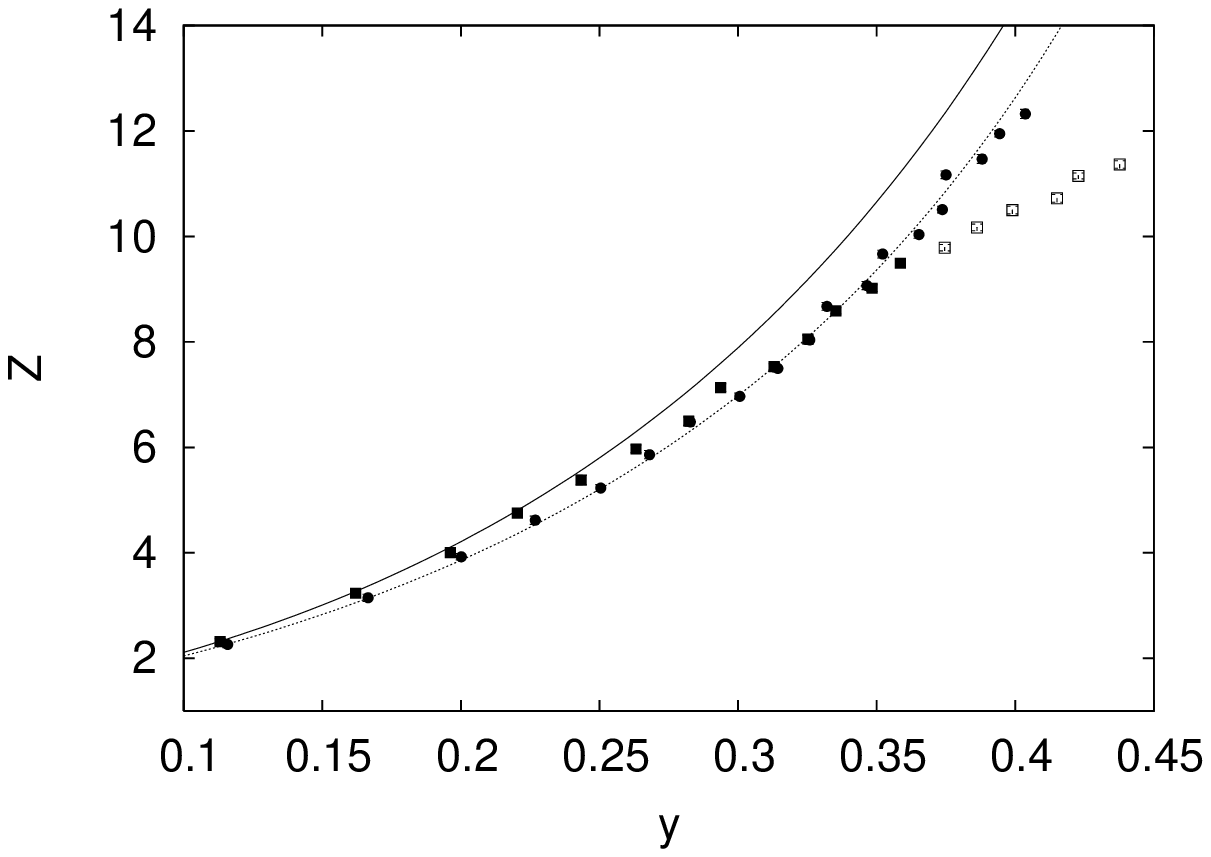}%
 \caption{\label{fig_5} Plot of the  results for the oblate $1\times5\times5$ model (isotropic: black squares, discotic: open squares) and the
prolate $1\times1\times5$ model (black circles) along with the Vega EOS (oblate: solid line, prolate: dashed line)}
 \end{figure}
 \begin{figure}
 \includegraphics[height=200pt,width=250pt]{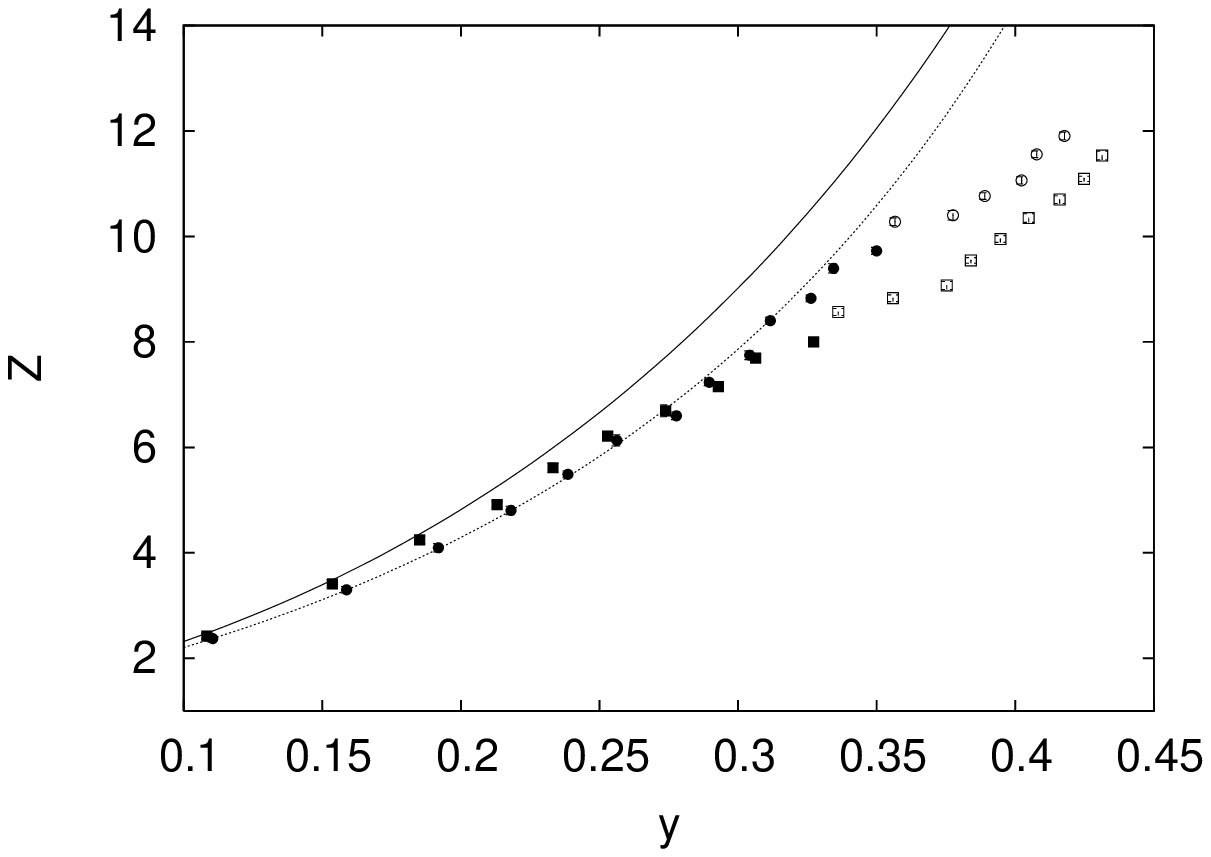}%
 \caption{\label{fig_6} Plot of isotropic the results for the oblate $1\times6\times6$ model (isotropic: black squares, discotic: open squares) and the
prolate $1\times1\times6$ model (isotropic: black circles, nematic: open circles) along with the Vega EOS (oblate: solid line, prolate: dashed line)}
 \end{figure}
 \begin{figure}
 \includegraphics[height=200pt,width=250pt]{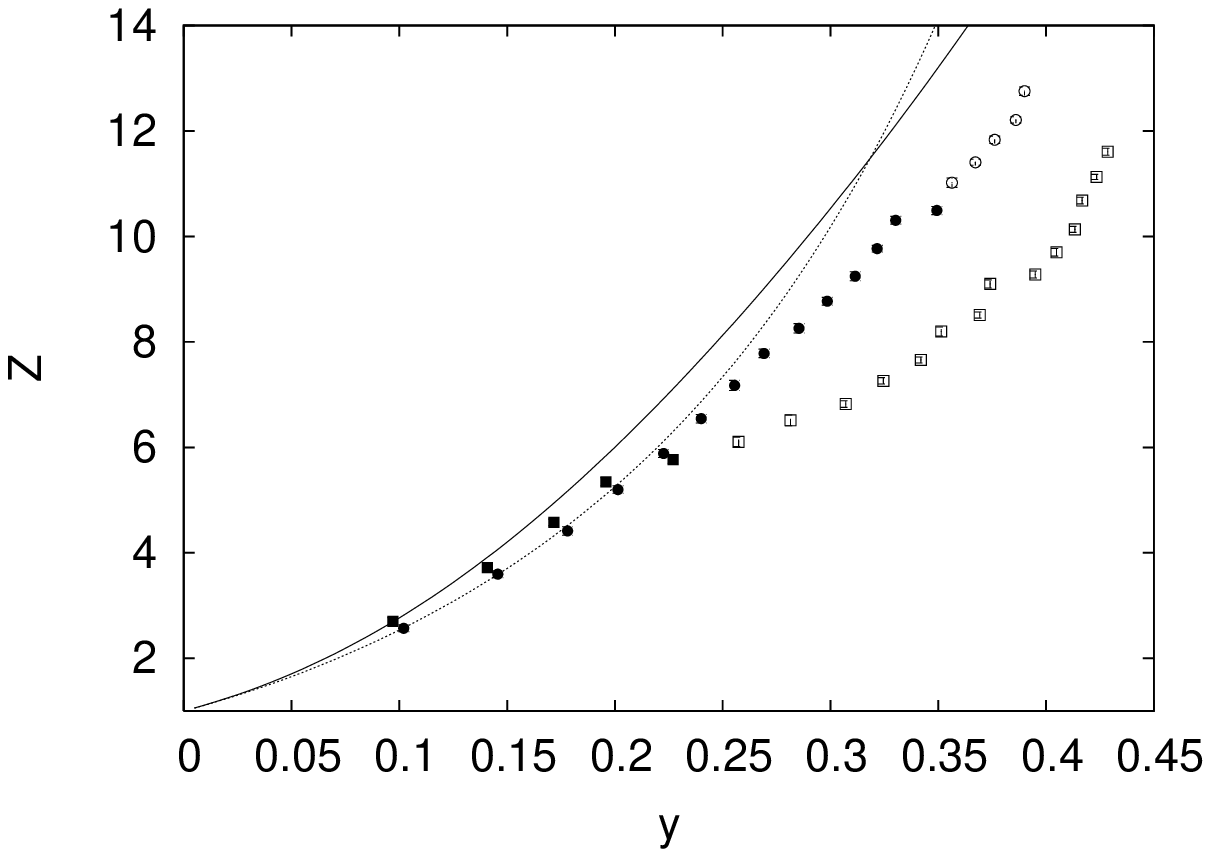}%
 \caption{\label{fig_8} Plot of isotropic the results for the oblate $1\times8\times8$ model (isotropic: black squares, discotic: open squares) and the
prolate $1\times1\times8$ model (isotropic: black circles, nematic: open circles) along with the Vega EOS (oblate: solid line, prolate: dashed line)}
 \end{figure}
 \begin{figure}
 \includegraphics[height=200pt,width=250pt]{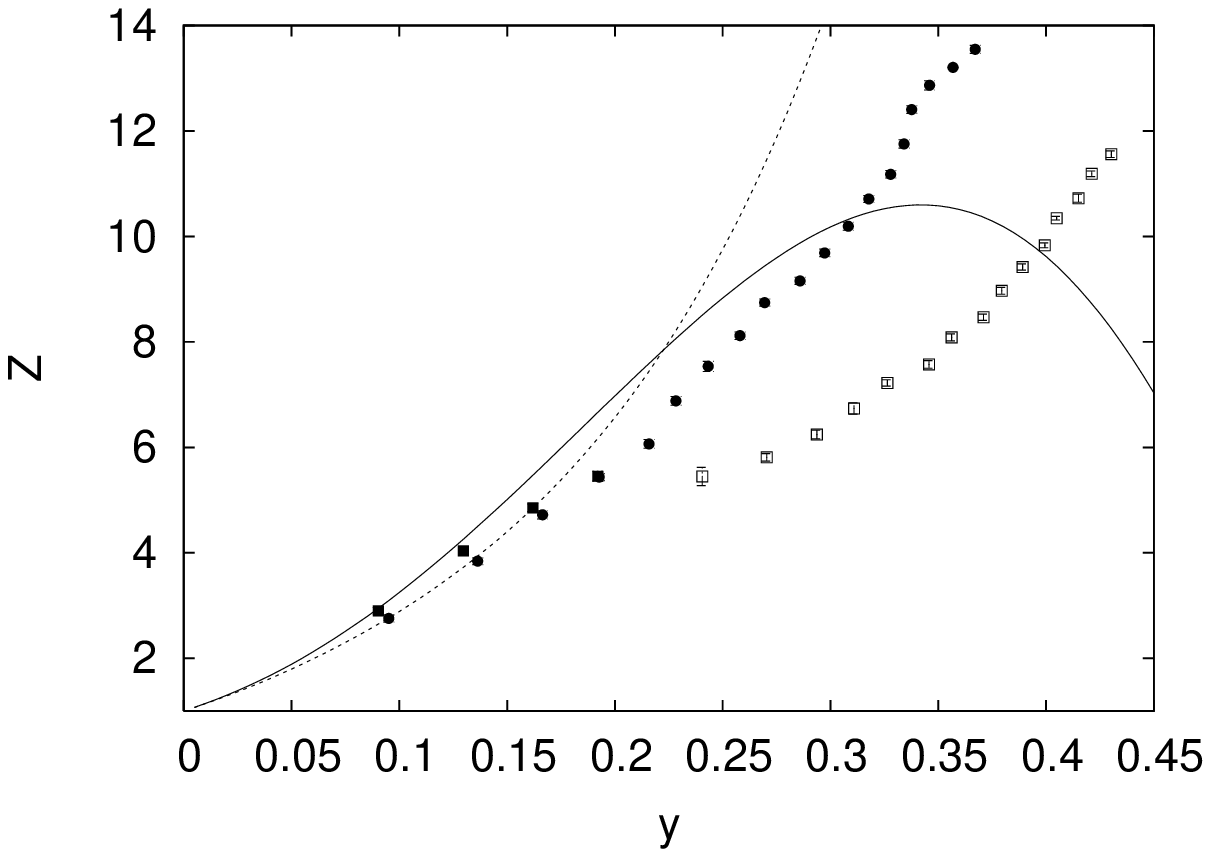}%
 \caption{\label{fig_10} Plot of the isotropic results for the oblate $1\times10\times10$ model (isotropic: black squares, discotic: open squares) and the
prolate $1\times1\times10$ model (black circles) along with the Vega EOS (oblate: solid line, prolate: dashed line)}
 \end{figure}
The theory behind the isotropic-nematic (I-N) transition was first developed 
by Lars Onsager \cite{ANYAS_1949_51_0627_nolotengo} and was studied for solutions of hard ellipsoids by Akira Isihara \cite{JCP_1951_19_01142}.
Frenkel and Mulder \cite{MP_1985_55_1171,MP_1985_55_1193} 
found an I-N$_+$ transition for a system of $\approx 90-108$  prolate ellipsoids of $c/a \geq 2.75$.
This result was called into question by Zarragoicoechea {\it et al.} \cite{MP_1992_75_0989},
suggesting system size effects play an important role in locating the I-N transition. However, a later work
by Allen and Mason \cite{MP_1995_86_0467} confirmed the nematic phase for $c/a = 3.0$ at densities of
$\rho/\rho_{cp}\approx 0.73-0.75$.
In this work no such transition is found until $c/a = 6$. 
It is interesting to compare this to linear tangent hard spheres, where
nematic phases were found for $m=6$ and a smectic A phase for $m=5$ where $m$ is the number of monomer
units \cite{JCP_2001_115_04203} (Note that smectic phases are not observed for hard ellipsoids).

The appearance of the nematic phase for the $1\times1\times8$ (Fig. \ref{fig_8})
model was considerably `delayed', 
and no nematic phase formed for compression runs (150k equilibration followed by 150k production) of the prolate $1\times1\times10$ model (Fig. \ref{fig_10}).
Given that it is fully expected to see $N_+$ phases for this model this
indicates that so called `jamming' is a real problem for especially elongated prolate systems.
The system finds itself grid-locked, and is unable to reorientate into 
the energetically more favourable nematic phase within the time scale of the simulation.
The equation of state of the glassy state (Fig. \ref{fig_10}) can be seen to be intermediate between the 
higher compressibility factor of the isotropic phase, indicated by the Vega EOS, and the much lower
values for the discotic branch of the $1\times10\times10$ model.
On the other hand, discotic phases readily appeared for the $1\times5\times5$ model (Fig. \ref{fig_5}) upwards.

Samborski {\it et al.} \cite{MP_1994_81_0263} placed the I-N transition  at 
$0.370 < y < 0.388$ for the $1\times1\times5$ model,
$0.333 < y < 0.351$ for the $1\times5\times5$ model,
$0.203 < y < 0.222$ for the $1\times1\times10$ model and
$0.185 < y < 0.203$ for the $1\times10\times10$ model.
It is interesting to note that I-N transition sets in at lower volume fractions for 
oblate ellipsoids than for prolate models that have the same value of $c$.
The same is seen in this work for the `6' (Fig. \ref{fig_6})  and `8' (Fig. \ref{fig_8})  models.
This is probably due to the fact that the compressibility factor of the 
oblate models, for a given volume fraction, is higher than that of its corresponding
prolate counterpart. This triggers the earlier onset of the I-N$_-$ transition.

\begin{figure}
 \includegraphics[height=200pt,width=250pt]{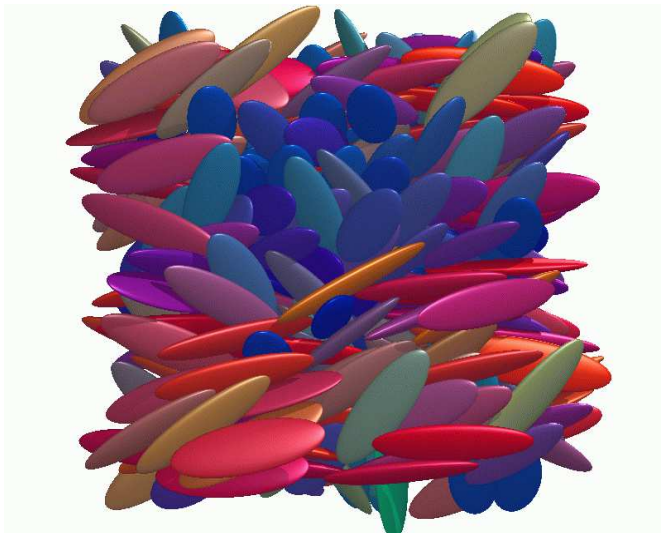}%
 \caption{\label{biaxial_photo} Snap-shot of the  $1\times3\times10$ model in the biaxial phase (here at $p^*=8.0$).
The ellipsoids are colour coded with respect to their orientations (colour online).}
 \end{figure}

\begin{figure}
 \includegraphics[height=200pt,width=250pt]{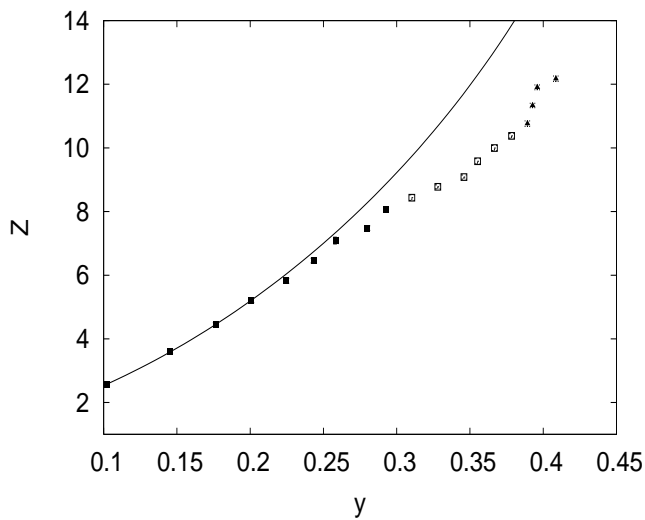}%
 \caption{\label{fig_1_3_10} Plot of the equation of state for the biaxial $1\times3\times10$ model (isotropic: black squares, discotic: open squares, biaxial: black triangles)
along with the Vega EOS for the isotropic branch.}
 \end{figure}

In this study only one convincing biaxial phase was identified, that of the  $1\times3\times10$
model (see Fig. \ref{biaxial_photo} for a snapshot, Fig. \ref{fig_1_3_10} for the EOS).
The biaxial phase formed at $p^*=8.0$, however, at $p^*=5.0$ the system had formed a discotic phase.
This indicates that the formation of a biaxial phase is a two-stage process; first orientating
along the short $a$ axis, later followed by the long $c$ axis {\it i.e.} 
having the phase transitions isotropic - discotic - biaxial.
This is in line with the observation that discotic phases form at lower volume fractions than nematic phases
(the biaxial phase can be seen as being composed of both a discotic and a nematic phase).

The positions of the isotropic liquid-liquid crystal transitions found in this work are presented
in Table \ref{i_n_table}.
A study of the I-N transition for biaxial ellipsoids was undertaken by Tjipto-Margo and Evans \cite{JCP_1991_94_04546}  
confirming the finding of Gelbart and Barboy \cite{ACR_1980_13_0290}  that biaxiality reduces the first order nature
of the I-N transition (this result was also confirmed for hard sphero-platelets by Somoza and Tarazona \cite{MP_1992_75_0017}).
It appears that the additional degree of freedom associated with biaxiality increases the disorder in the 
nematic phase with respect to a uniaxial ($D_{\infty h}$) model.       
At the `self-dual' point (where $a:b=b:c$,  i.e. $b=\sqrt{ac}$) the transition becomes second-order \cite{PRA_1989_39_000360}.
The isotropic-biaxial nematic transition is said to occur for models that are close to this self-dual point \cite{LC_1990_8_0499}.
For the uniaxial $1\times10\times10$ model a considerable jump can be seen in the volume fractions associated 
with the formation of the discotic phase (Fig. \ref{fig_10}), one of the hall-marks of a 
first-order transition. Meanwhile, for the biaxial $1\times3\times10$ model no such jump is seen.

\begin{table}%[H] add [H] placement to break table across pages
\caption{Location of the isotropic-nematic transition ($S_2 > 0.4$) for hard ellipsoids (Note: $a=1$).
Order parameters are given for $p^*=9.5$.}
\label{i_n_table}
%\begin{ruledtabular}
%~~~~~~~$b$~~~~~~~~~$c$~~~~~~~~$p*$~~~~~~~~~~~$y$\\
 \begin{ruledtabular}
\begin{tabular}{|r|r|r|r|r|r|r|}
\hline
$b$ & $c$ & $p^*$ & $y$ & phase & $S_2(c)$ & $S_2(a)$\\
\hline
prolate \\
\hline
1 & 2.5 & --     &  --         & I     & $< 0.1$  & $< 0.1$ \\
1 & 4 &   --     &  --         & I     & $< 0.1$  & $< 0.1$  \\
1 & 5 &   --     &  --         & I     & 0.13     &  $< 0.1$  \\
1 & 6 &   7.0    &  0.356    & $N_+$   & 0.69 & 0.20 \\
1 & 8 &   7.5    &  0.356         & $N_+$ & 0.55 & 0.18 \\
1 & 10 &  --     &  --         & {\it I} & 0.30 & 0.12 \\
\hline
oblate\\
\hline
2.5 & 2.5 &  --     &  --       & I    & $< 0.1$  &  $< 0.1$  \\
4 & 4 &      --  &     -- & I          & 0.12 & 0.24 \\
5 & 5 &      7.0    &     0.374 & $N_-$& 0.23 & 0.81 \\
6 & 6 &      5.5    &     0.336 & $N_-$& 0.25 & 0.87 \\
8 & 8 &      3.0    &     0.257 & $N_-$& 0.26 & 0.92 \\
10 & 10 &    2.5    &     0.240 & $N_-$& 0.27 & 0.95 \\
\hline
biaxial prolate & ($b < \sqrt{ac} $)\\
\hline
2 & 5 &  --     &  --         & I      & 0.10 & $< 0.1$ \\
2 & 6 &  --     &  --         & I      & 0.21 &  0.15\\
2 & 8 &  6.0    &  0.342 &$N_+$        & 0.85 & 0.25 \\
3 & 10 & 8.0    &  0.389 &$B$        & 0.48 & 0.76 \\
\hline
biaxial self-dual\\
\hline
1.25 & 1.5625 &  --     &  --     & I  & $< 0.1$ &  $< 0.1$\\
2 & 4 &  --     &  --         & I      & $< 0.1$ &  $< 0.1$\\
3 & 9 &  4.5     & 0.30  & $N_-$       & 0.34    &  0.81   \\
\hline
biaxial oblate  & ($b > \sqrt{ac} $)\\
\hline
3 & 6 &  9.0    &  0.402 &$N_-$        & 0.23 & 0.46 \\
3 & 8 &  6.0    &  0.338 &$N_-$        & 0.27 & 0.78 \\
5 & 8  & 4.5    &  0.313 &$N_-$        & 0.25 & 0.90 \\ 
5 & 10 & 3.5    &  0.269 &$N_-$        & 0.30 & 0.92 \\
8 & 10 & 3.0    &  0.249 &$N_-$        & 0.27 & 0.95 \\
\hline
 \end{tabular}
 \end{ruledtabular}
 \end{table}
%%%%%%%%%%%%%%%%%%%%%%%%%%%%%%%%%%%%%%%%%%%%%%%%%%%%%%%%%%%%%%%%%%%
%\subsection{density change across the transition}
%%%%%%%%%%%%%%%%%%%%%%%%%%%%%%%%%%%%%%%%%%%%%%%%%%%%%%%%%%%%%%%%%%%

%%%%%%%%%%%%%%%%%%%%%%%%%%%%%%%%%%%%%%%%%%%%%%%%%%%%%%%%%%%%%%%%%%%
\section{Conclusions}
%%%%%%%%%%%%%%%%%%%%%%%%%%%%%%%%%%%%%%%%%%%%%%%%%%%%%%%%%%%%%%%%%%%
Various ellipsoidal models have been subjected to Monte Carlo compression runs.
The Vega equation of state is seen to perform very well for the isotropic phases
of all of the models considered. 
The elongated uniaxial models show indications of the first-order nature of the 
isotropic-nematic transition. However, the formation of orientationally ordered phases for 
very long prolate models is severely hindered by the formation 
of a glass like state. The oblate models form discotic readily, at lower volume fractions than 
their prolate partners.
The biaxial phase seems to form in a two-stage process, first forming 
a discotic phase, followed by the biaxial phase upon orientation of the 
long axes.

\begin{acknowledgments}
The authors should like to thank
C. Vega for the provision of the Perram-Wertheim overlap
algorithm and N. G. Almarza for useful discussions.
This work was funded by project FIS2004-02954-C03-02 of the Spanish
Ministerio de Educacion y Ciencia, and by project S-0505/ESP/0299 - CSICQFT
(MOSSNOHO) of the D. G. de Universidades e Investigación del Comunidad de Madrid.
One of the authors (C. M.) would like to thank the  CSIC for the award of an I3P post-doctoral contract.
\end{acknowledgments}
%%%%%%%%%%%%%%%%%%%%%%%%%%%%%%%%%%%%%%%%%%%%%%%%%%%%%%%%%%%%%%%%%%%
\section{Appendix A}
%%%%%%%%%%%%%%%%%%%%%%%%%%%%%%%%%%%%%%%%%%%%%%%%%%%%%%%%%%%%%%%%%%%
 \begin{table}%[H] add [H] placement to break table across pages
%\caption{\label{}}
%\begin{tabular}{}
\caption{Values for $R$, $S$, $V$, $\alpha$ and $B_2/V$ for hard ellipsoids (Note: $a=1$).}
\label{rsv_table}
%\begin{ruledtabular}
%~~~~~~~$b$~~~~~~~~~$c$~~~~~~~~$R$~~~~~~~~~~~$S$~~~~~~~~~~~$V$ ~~~~~~~~~$\alpha$ ~~~~$B_2/V$\\
 \begin{ruledtabular}
\begin{tabular}{|r|r|r|r|r|r|r|}
\hline
$b$ & $c$ & $R$ & $S$ & $V$  & $\alpha$ & $B_2/V$\\
\hline
sphere \\
\hline
1 &1    &         1&  $4\pi$  &  $4\pi/3$ & 1 & 4\\
\hline
prolate \\
\hline
1 & 2.5 & 1.5919      & 26.1518    &  10.472    &  1.32516 & 4.97549\\
1 & 4 &   2.26639    &  40.4975   &   $16\pi/3$ &  1.82597 & 6.4779 \\
1 & 5 &   2.73397   &   50.1925   &   $20\pi/3$ &  2.184   & 7.552 \\
1 & 6 &   3.20942    &  59.9386   &   $8\pi$    &  2.55136 & 8.65409\\
1 & 8 &   4.17441    &  79.5147   &   $32\pi/3$ &  3.30174 & 10.9052\\
1 & 10 &  5.15042     & 99.151    &   $40\pi/3$ &  4.06377 & 13.1913\\
\hline
oblate\\
\hline
2.5 & 2.5 &  2.0811     &  50.0111   & 26.1799 &  1.32516  & 4.97549 \\
4 & 4 &      3.22269 &    113.921 &  $64\pi/3$ &  1.82597  & 6.4779\\
5 & 5 &      3.99419 &  171.78  &   $100\pi/3$ &  2.184  & 7.552 \\
6 & 6 &      4.76976 &    241.985 &  $48\pi$   &  2.55136  & 8.65409 \\
8 & 8 &      6.32758 &    419.657 &  $256\pi/3$ & 3.30174  & 10.9052 \\
10 & 10 &    7.89019    & 647.22  &  $400\pi/3$ & 4.06377  & 13.1913 \\
\hline
biaxial\\
\hline
1.25 & 1.5625 &  1.28328     & 20.1576   & 8.18123  &  1.05395   &  4.16185 \\
2 & 4 &  2.52566     &   63.4766  &  $32\pi/3$ &  1.59473   &  5.7842 \\
2 & 5 &  2.96925     &   78.2743 &  $40\pi/3$ &   1.84951  &   6.54853\\
2 & 6 &  3.42527     &   93.1895   & $16\pi$   &  2.11675    & 7.35026 \\
2 & 8 &  4.36059     &  123.218   &  $64\pi/3$ &  2.67234  &  9.01701  \\
3 & 6 &  3.70789     &  129.13   & $24\pi$   &   2.11675   &  7.35026 \\
3 & 8 &  4.60996     &   170.448  & $32\pi$   &  2.60536   &  8.81607  \\
3 & 9 &  5.07182     &   191.203  & $36\pi $   & 2.85815     & 9.57446 \\
3 & 10 & 5.53883      & 212.0    & $40\pi$   &   3.11474  &  10.3442 \\
5 & 8 &  5.22725     &  268.73   &  $160\pi/3$ & 2.7946    &  9.38379   \\
5 & 10 & 6.10332      & 333.946    & $200\pi/3$   &  3.24386   & 10.7316 \\
8 & 10 & 7.1304      &   521.211  &  $320\pi/3$   &  3.69682   & 12.0904 \\
\hline
 \end{tabular}
 \end{ruledtabular}
 \end{table}
In Table \ref{rsv_table} we provide a table of $R$, $S$, $V$, $\alpha$ and $B_2/V$ for the various models
studied in this work.
The values for  $R$ and $S$ are obtained by evaluating the 
expressions derived by Singh and Kumar \cite{JCP_1996_105_02429,AoP_2001_294_0024}.
Thus the mean radius of curvature is given by
\begin{equation}
R= \frac{a}{2} \left[  \sqrt{\frac{1+\epsilon_b}{1+\epsilon_c}} + \sqrt \epsilon_c \left\{ \frac{1}{\epsilon_c} F(\varphi , k_1) + E(\varphi,k_1) \right\}\right],
\end{equation}
and the surface area by
\begin{equation}
S= 2 \pi a^2 \left[  1+  \sqrt {\epsilon_c(1+\epsilon_b)} \left\{ \frac{1}{\epsilon_c} F(\varphi , k_2) + E(\varphi,k_2)\right\} \right],
\end{equation}
where $F(\varphi,k)$ is an elliptic integral of the first kind and $E(\varphi,k)$ is an elliptic integral of the second kind,
with the amplitude being
\begin{equation}
\varphi = \tan^{-1} (\sqrt \epsilon_c),
\end{equation}
and the moduli
\begin{equation}
k_1= \sqrt{\frac{\epsilon_c-\epsilon_b}{\epsilon_c}},
\end{equation}
and
\begin{equation}
k_2= \sqrt{\frac{\epsilon_b (1+\epsilon_c)}{\epsilon_c(1+\epsilon_b)}},
\end{equation}
where the anisotropy parameters, $\epsilon_b$ and $\epsilon_c$,  are
\begin{equation}
\epsilon_b = \left( \frac{b}{a} \right)^2 -1,
\end{equation}
and
\begin{equation}
\epsilon_c = \left( \frac{c}{a} \right)^2 -1.
\end{equation}
The volume of the ellipsoid is given by the well known
\begin{equation}
V = \frac{4 \pi}{3}abc.
\end{equation}
Note the symmetry between the prolate and oblate ellipsoids for the values of $\alpha$,
and thus for $B_2$. 
\bibliography{bibliography}

\end{document}